\documentclass[prd,aps,preprint]{revtex4}

\usepackage{epsfig}
\usepackage{subfigure}
\usepackage{color}
\usepackage{array}
\usepackage{url}
\input epsf.tex
\def\DESepsf(#1 width #2){\epsfxsize=#2 \epsfbox{#1}}
\begin{document}

%\draft

%@twocolumnfalse\endcsname
\preprint{\vbox{
\hbox{MIFP-09-40,
UMD-PP-09-050} }}

\title{{\Large\bf Origin of Quark-Lepton Flavor
in SO(10) with Type II Seesaw }}
\author{{\bf Bhaskar Dutta}$^1$, {\bf Yukihiro Mimura}$^2$ and {\bf R.N.
Mohapatra}$^2$ }

\address{
$^1$Department of Physics, Texas A\&M University,
College Station, TX 77843-4242, USA
\\
$^2$ Maryland Center for Fundamental Physics and Department of Physics,
University of Maryland, College Park, MD, 20742, USA}
\date{October, 2009}
%\maketitle

\begin{abstract}
Diverse mass and mixing patterns between the quarks and leptons
makes it challenging to construct a simple grand unified theory of
flavor. We show that SO(10) SUSY GUTs with type II seesaw
mechanism giving neutrino masses provide a natural framework for
addressing this issue. A simple ansatz that the dominant Yukawa
matrix (the {\bf 10}-Higgs coupling to matter) has rank one,
appears to simultaneously explain both the large lepton mixings as
well as the observed quark flavor hierarchy in these models. A testable
prediction of this ansatz is the neutrino mixing, $U_{e3}$, which should
be observable in planned long baseline experiments.
 \end{abstract}

\maketitle

\vskip1.0in

\newpage

\baselineskip 18pt

\section{Introduction}
Understanding the origin of the hierarchical pattern of quark masses and
mixings has long been recognized as a challenge for physics beyond the
standard model \cite{weinberg}. The discovery of neutrino masses and
mixings with totally
different flavor pattern than quarks (i.e. $\theta^l_{23}\sim 45^{\rm o}$ and
$\theta^l_{12}\simeq 35^{\rm o}$ as against $\theta^q_{23}\sim 2.5^{\rm o}$ and
$\theta_{12}^q\sim 13^{\rm o}$) has added more mystery to the flavor problem.
In generic bottom-up pictures where quarks and leptons are treated as
different species of particles with no particular relation between them,
this problem is not so serious since one can simply focus on each sector
separately, as is often done for neutrinos \cite{review}. However, in
grand unified theories where the quarks
and leptons unify at a very high scale, one would naively expect that
their masses and mixings would exhibit a similar pattern. The fact that they
are so different may be hint of some really new exciting underlying
physics. In this note we address this question in the context of
supersymmetric SO(10) models with renormalizable Yukawa couplings being
responsible for fermion masses.

We show that in SO(10) models with {\bf 10}, {\bf 126} plus
possibly another {\bf 10} or {\bf 120} Higgs fields where fermion
masses are generated by renormalizable Yukawa couplings \cite{babu}
only and where type II seesaw is responsible for neutrino
masses \cite{goran}, there is a natural way to have a unified
understanding of both large lepton mixings and small quark ones.
The basic idea is to require that one of the {\bf 10} Yukawa couplings
is the dominant one contributing to up, down and charged lepton masses
and has rank one with other smaller couplings providing neutrino
masses as well as most of the quark lepton flavor hierarchy.
Rank one plus small corrections as a way to unravel fermion
flavor in D-brane models was discussed in \cite{dutta}. We find that
SO(10) models with type II seesaw \cite{babu,goran,so10,mimura} are
ideally suited for such
an ansatz. A specific form of the rank one matrix can lead to
tri-bimaximal mixing with corrections dictated by the quark flavor
pattern.

This paper is organized as follows: in sec. II, we review the mass
formulae in SO(10) models with renormalizable couplings; in sec. III, we
summarize our basic strategy for understanding the quark lepton flavor in
a unified manner, discuss the rank one ansatz and apply it both the two
generation case (IIIA) and three generation cases (IIIB). In sec. IIIC,
we present realistic
three generation models and outline their predictions. Sec. IV is devoted
to some specific conjectures for the rank one matrix which could emerge
from discrete symmetry models with specific discussion on the correction
to tri-bimaximal mixings. Sec. V is devoted to a possible way to obtain
the rank one ansatz in SO(10) models and in sec. VI, we present our
conclusions.

\section{Overview of renormalizable SUSY SO(10) models for fermion masses}
The basic idea in this class of models is to consider SUSY SO(10) theory
with Higgs fields that
give fermion masses to be in {\bf 10} (denoted by $H$) and  ${\bf
126}+\overline{\bf 126}$
(denoted by $\Delta$ and $\overline{\Delta}$) plus either an extra
{\bf 10} ($H'$) or {\bf 120} ($\Sigma$). The GUT symmetry
is broken by {\bf
210}+{\bf 54}+{\bf 126}+$\overline{\bf 126}$ \cite{breaking}.  The Yukawa
superpotential of this model is:
\begin{eqnarray}
W_Y~=~h\, \psi\psi H + f\, \psi\psi\bar{\Delta}+h'\,\psi\psi \, (\Sigma \ {\rm or} \
H')\,,
\end{eqnarray}
where the symbol $\psi$ stands for the {\bf 16} dimensional
representation of SO(10) that represents the matter fields.
The coupling matrices $h$ and $f$ are symmetric,
and $h^\prime$ is symmetric or anti-symmetric depending on whether we
adopt $H^\prime$ or  $\Sigma$.
The representations $H$, $H'$ and $\Delta$ have two standard model
(SM) doublets in each of them whereas $\Sigma$ has four such doublets.
The general
way to understand so many SM doublets is that at the GUT scale $M_U$,
once the GUT and the $B-L$ symmetry are broken, one linear
combination of the up-type doublets and one of down-type ones
remain almost massless whereas the remaining ones acquire GUT
scale masses just like the color triplet and other non-MSSM
multiplets.
The electroweak symmetry is broken after the light MSSM doublets
(to be called $H_{u,d}$) acquire vacuum expectation values (vevs) and
they then generate the
fermion masses. The resulting mass formulae for  different
fermion masses are given by:
\begin{eqnarray}
Y_u &=& h + r_2 f +r_3 h^\prime, \label{eq2} \\\nonumber
Y_d &=& r_1 (h+ f + h^\prime)\,, \\\nonumber
Y_e &=& r_1 (h-3f + c_e h^\prime)\,, \\\nonumber
Y_{\nu^D} &=& h-3 r_2 f + c_\nu h^\prime,
\end{eqnarray}
where $Y_a$ are mass matrices divided by the electro-weak vev
$v_{wk}$ and $r_i$ and $c_{e,\nu}$ are the mixing parameters which
relate the $H_{u,d}$ to the doublets in the various GUT
multiplets.
More precisely,
the matrices $h$, $f$ and $h^\prime$ in $Y_a$
are multiplied by the Higgs mixings.
The precise definitions of the couplings and the Higgs mixings are given
in ref.\cite{mimura}.
When $H^\prime$ is adopted for the $h^\prime$ coupling,
$c_e= 1$ and $c_\nu = r_3$.
In generic SO(10) models of this type, the neutrino
mass formula has a type I \cite{type1} and a type II \cite{type2}
contributions:
\begin{eqnarray}
{\cal M}_\nu~=~fv_L-M_D\frac{1}{fv_R}M^t_D\,,
\end{eqnarray}
where $v_L$ is the vev of the $B-L=2$ triplet in the {\bf 126} Higgs
field and is given by $v_L\simeq \frac{\lambda \mu
v^2_{wk}}{M^2_{\Delta_L}}$. Note that in general, the two
contributions to neutrino mass depend on two different parameters
and it is easy to have symmetry breaking pattern in
SO(10) \cite{goh} where the first contribution (the type II term)
dominates over the type I term. The neutrino mass formula then
becomes
\begin{eqnarray}
{\cal M}_\nu~=~fv_L.
\label{eq4}
\end{eqnarray}
Note that $f$ is the same coupling matrix that appears in the
charged fermion masses in Eq. (\ref{eq2}),
up to factors from the Higgs mixings and the Clebsch-Gordan coefficients.
The equations (\ref{eq2}) and (\ref{eq4}) are
the key equations in our unified approach to address the flavor problem.

The main hypothesis of our approach is that the fermion mass
formula of Eq. (\ref{eq2}) are dominated by the matrix $h$ with the
contributions of $f$ and $h'$ being small perturbations. In the
limit of $f,h'\to 0$, the quark and lepton mixings vanish as do
the neutrino masses. We will show below that this simple hypothesis
combined with Eq. (\ref{eq4}) can simultaneously explain large lepton
mixings while keeping the quark mixings being proportional to
$|f|/|h|$ and hence small. We will subsequently assume that the
matrix $h$ has rank one in which case the mass hierarchy can also be
explained in a natural manner.

\section{Explaining quark-lepton flavor hierarchies}

The quark and lepton mixing matrices are given by
the product of diagonalizing unitary matrices for quark and lepton mass
matrices as follows: denoting the diagonalizing matrices of $M_u$, $M_d$
by $V_u$ and $V_d$ respectively (e.g., $V_u M_u M_u^\dagger V_u^\dagger =
{\rm diag.}(m_u^2, m_c^2, m_t^2)$ and similarly for the down quark mass
matrix ),
the CKM (Cabibbo-Kobayashi-Maskawa) quark mixing matrix is given as
$V_{\rm CKM} = V_u V_d^\dagger$.
The PMNS (Pontecorvo-Maki-Nakagawa-Sakata) lepton mixing matrix is given
as $U_{\rm PMNS} = (V_e V_\nu^\dagger)^*$
in the similar notation (e.g., $V_\nu {\cal M}_\nu V_\nu^t = {\rm
diag.}(m_1,m_2,m_3)$).

In general, when two matrices with random $O(1)$ elements are considered,
the mixing angles of the relative diagonalizing unitary matrices
are all $O(1)$ in radian,
while the eigenvalues can have a hierarchy of $O(0.1)$.
In such an anarchical scenario, the neutrino masses and mixings can be
explained
(except for the CHOOZ bound of 13 neutrino mixing):
the neutrino mixings are generically $O(1)$ and there is a little
hierarchy for
the neutrino mass squared difference ratio
$\Delta m^2_{12}/\Delta m^2_{23}$ \cite{Haba:2000be}.
On the other hand, since the quark mixings are all smaller than $O(1)$
and the masses of quarks and charged leptons are very much hierarchical,
anarchic mass matrices in general provide no explanation of these
observations.
Besides, it appears that the mass ratios and CKM mixings have several
correlations among them. It is therefore to be
 expected that the quark and lepton matrices instead of being independent
anarchic matrices must have some relations among them
and an underlying theory leading to these relations.
 In this paper we find that SO(10) with type II seesaw could be such a
theory.

When the fermion masses are given by the Eqs. (\ref{eq2}) and (\ref{eq4}),
several possible outcomes are obtained by simple assumptions.
To understand these possible outcomes from the Eqs.(\ref{eq2}) and
(\ref{eq4}), let us first
ignore $h^\prime$.
We then have the following possibilities:

\bigskip

\noindent{\it Assumption 1:}

Take $h$, $f$ are general rank 3 matrices, and $f$ is small. This is the case
analyzed to fit observed experimental data and to obtain predictions
from the minimality of the number of parameters
in various papers \cite{goran}.
Here, we list the properties resulting from the smallness of $f$ without
resorting to any numerical fit.

\begin{itemize}
    \item The CKM mixings are small, due to the fact that there is an
approximate up-down symmetry and $V_{\rm
CKM}=V_uV_d^\dagger$ \cite{babu1}.
    \item Bottom-tau unification up to $O(f/h)$.
    \item The 3 neutrino mixings are generically of $O(1)$ since $h$ and
$f$ are unrelated matrices. The type II seesaw dominance of the neutrino
mass is crucial for the generic largeness of the neutrino mixings.
\end{itemize}
Thus it is interesting that without any special assumption, the gross
features of fermion mixings can be reproduced.
%Note that the reason why the neutrino mixings are generically large
%is because we assume that the type II term dominates the neutrino mass.
This does not,
however, throw any light on the mass hierarchies among quarks and
leptons, though one can fit the experimental results by the choices of
parameters
(even in the type I seesaw) \cite{goran,Matsuda:2000zp,Babu:2005ia}.
Since we use the experimental data as an input, these scenarios do not
provide a fundamental understanding of
either the mass hierarchy for quarks and charged leptons,
or why the 13 neutrino mixing ($U_{e3}$) is less than $O(1)$.
 Similar situation holds for models where $h^\prime$ is
added \cite{Dutta:2004wv,others}.

\bigskip

\noindent{\it Assumption 2:}

Let us next consider the specific case when $h$ is a rank 1
matrix \cite{mimura}, and $f$ is a rank 3 matrix with
 eigenvalues of $f$ being hierarchical ($f_1,f_2 \ll f_3$) and small
compared to the elements of $h$. As we noted in ref.\cite{mimura},
this choice helps to suppress proton decay
 in SUSY SO(10) models without invoking huge cancellations among the
colored Higgsino exchange amplitudes. In this
 case we will show in the next two subsections that the following results
follow:

\begin{enumerate}
    \item CKM mixings are small.
    \item Approximate bottom-tau unification occurs.
    \item %We find
           $\displaystyle \frac{m_c}{m_t} : \frac{m_s}{m_b} :
\frac{m_\mu}{m_\tau}
\simeq r_2 : 1: -3$.
    \item  The quark mixing are related as $V_{cb} \sim m_s/m_b +
e^{i\sigma} m_c/m_t$ (where $\sigma$ is a phase)
and $V_{ub} \sim V_{cb} f_2/f_3$.
    \item Atmospheric and solar neutrino mixings are generically large,
but 13 mixing is $\sim f_2/f_3$.
\end{enumerate}

All these predictions are in qualitative agreement with observations.
The advantage of the rank one assumption is that it naturally explains the
mass hierarchies among quarks and leptons in addition to large
lepton mixings.
We emphasize that these features are obtained from the rank 1 assumption above
without using any numerical inputs,
and our claim here is not based on the scenario of the numerical
predictions from a fit in which the minimality of  parameters plays a key
role.

Before we do a full demonstration of these results in the context of
a three generation model, let us illustrate the first four points
in the context of a two generation model.
% The three generation fit
%requires a non-zero $h^\prime$, as we see below.

\subsection{A two generation illustration}
In this subsection, we apply our rank one hypothesis to the second and the
third generation. We will confirm the
results 1-4 mentioned above. The starting point is the mass relation
from Eqs. (\ref{eq2}) and (\ref{eq4}) where we ignore the $h^\prime$
contribution .
%to Yukawa matrices (ignoring $h^\prime$):
%\begin{eqnarray}
%Y_u &=& h + r_2 f \\
%Y_d &=& r_1 (h+ f ) \\
%Y_e &=& r_1 (h-3f ) \\
%{\cal M}_\nu~=~fv_L
%\end{eqnarray}
Using our assumption,
we have
$h=\begin{pmatrix}{ \sin\theta & \cos\theta} \end{pmatrix}^t
\begin{pmatrix}{\sin\theta & \cos\theta}\end{pmatrix} h_3$
and $f={\rm diag}(f_2, f_3)$
(without loss of generality, we can parameterize $f$ to be diagonal).
The parameter $\theta$ is of $O(1)$ in general.
We now have ten parameters ($\theta$, $h_3$ and $r_1$ as real parameters,
$f_2$, $f_3$, and $r_2$ as complex parameters, and $v_L$ for neutrino
mass scale)
describing ten observables of all lepton and quark mixings and
masses.

%In this rotated basis, the $b$-quark mass is given by $m_b\sim r_1 m_t$,
%which therefore fixes $r_1$ to be about $\sim 1/50$ and also we have the
%relation $m_b\simeq m_\tau$ to the leading order in $\epsilon$.
One can easily obtain $r_1 m_t \simeq m_b \tan\beta \simeq m_\tau \tan\beta$
at the leading order neglecting $O(f_3/h_3)$ correction,
where $\tan\beta$ is a ratio of vevs of $H_u$ and $H_d$.
Therefore, $r_1$ corresponds to the freedom of $\tan\beta$,
$r_1 \sim \tan\beta/50$.

When $f_2 \ll f_3 \ll h_3$, we obtain
\begin{equation}
\frac{m_c}{m_t} \simeq r_2 \frac{f_3}{h_3} \sin^2\theta,
\quad
\frac{m_s}{m_b} \simeq \frac{f_3}{h_3} \sin^2\theta,
\quad
\frac{m_\mu}{m_\tau} \simeq -3 \frac{f_3}{h_3} \sin^2\theta.
\end{equation}
Because $m_c/m_t \ll m_s/m_b$, $r_2$ is small, i.e. $r_2 \simeq
m_c/m_t/(m_s/m_b)$.

To proceed further, we first diagonalize the charged fermion mass matrices
to zeroth order in $f_2,f_3 \to 0$.
The matrix diagonalizing this is given by
\begin{eqnarray}
U_0~=~\begin{pmatrix} {\cos\theta & -\sin\theta \cr \sin\theta &
\cos\theta}\end{pmatrix},
\end{eqnarray}
since $U_0 (\sin\theta \ \cos\theta)^t = (0\ 1)^t$.
Let us now see how the small quark mixings arise despite large mixings in
$U_0$.
Because $U_0 Y_d U_0^t$ has a small off-diagonal element,
$r_1 (f_2 - f_3) \sin \theta \cos \theta$,
the down-type quark mass matrix is diagonalized by
$V_d = \tilde V_d U_0$,
where $\tilde V_d$ is close to unit matrix whose off-diagonal element is
$\simeq f_3/h_3 \sin\theta \cos\theta$.
The up-type quark mass matrix is diagonalized by $V_u = \tilde V_u U_0$,
where the off-diagonal element of $\tilde V_u$ is $\simeq r_2 f_3/h_3
\sin\theta\cos\theta$.
The quark mixing matrix is then given by
$V_{\rm CKM} = V_u V_d^\dagger = \tilde V_u \tilde V_d^\dagger$,
in this product $U_0$
 cancels out leaving small mixings between the two generations i.e.
small $V_{cb} \simeq (1-r_2) f_3/h_3 \sin\theta\cos\theta
\simeq (m_s/m_b + e^{i\sigma} m_c/m_t) \cot\theta$,
where $\sigma$ is a phase of $r_2$.

%
%the up quark mass matrix to
%zeroth order in $r_2$ since $r_2\sim m_c/m_t\sim 140$, which is small
%compared to all the other parameters. The matrix diagonalizing this is
%given by
%\begin{eqnarray}
%U^0_L~=~\begin{pmatrix} {cos\theta & sin\theta \cr -sin\theta &
%cos\theta}\end{pmatrix}
%\end{eqnarray}
%Let us now see how the small quark mixings arise despite large mixings in
%$U^0_L$. To get the analog of CKM matrix, we neglect $m_c/m_t\sim 140$
%and see that $V_u= U^0_L$ diagonalyzes the up quark mass matrix.
%The down quark mass matrix is now diagonalized by $V_d~=~
%U^0_L\tilde{V}_d$ where
%$\tilde{V_d}$ is close to a
%unit matrix ( i.e. has small off diagonal elements).
% The quark mixing matrix is then given by
%$V_{CKM}=V^\dagger_dV_u$ and in this product $U^L_0$
% cancels out leaving small mixings between the two generations i.e.
%small $V_{cb}$.

Coming now to lepton mixings, suppose that the
charged lepton mass matrix is diagonalized by $V_\ell$, then it can be
written as $V_\ell = \tilde{V}_e U_0$,
where $\tilde V_e$ is close to a unit matrix
similarly to the quark sector, and is roughly equal to $V_{\rm CKM}^\dagger$.
%
%since the dominant contribution to it is from $h$ which is diagonalized by $U^0_L$.
%This then induces off-diagonal elements coming from the small $f$
%and $h'$ couplings which are diagonalized subsequently by $\tilde{V}_e$
%which is roughly equal to $V_{CKM}$.
%
Since the neutrino mass matrix is already diagonal as a parameterization,
the PMNS matrix is given by the charged lepton mixings so that
$U_{\rm PMNS}= V_\ell^* ~\simeq~V^{t}_{\rm CKM}U^0_L$.  This leads to a
large lepton mixing  as desired. In the two generation
case, $\theta$ describes approximately (up to small corrections of order
$V_{cb}$) the ``atmospheric mixing angle". Since this was an input into
our rank one ansatz, we can choose to be large to explain the
observations.
% In this case, we also see that
%$\frac{\epsilon_2}{\epsilon_3}+\frac{m_{\odot}{m_{atm}}$.

If $f_2,f_3$ and $r_2$ are assumed to be real,
there are six real parameters in this model.
In this case, $m_b, m_s$, $\theta_{\rm atm}$ and $m_2/m_3$
can be written as a function of $m_c, m_t$, $m_\mu, m_\tau$ and $V_{cb}$
for example.
Even if $f_2,f_3$ and $r_2$ are all complex,
we have the following approximate relation at the grand unified scale:
\begin{eqnarray}
\frac{m_s}{m_b}~= V_{cb} \tan\theta
\left(1+O\left(\frac{f_3}{h_3}\right)\right),  \\ \nonumber
m_b~= m_\tau \left(1+O\left(\frac{f_3}{h_3}\right)\right), \\ \nonumber
m_s~= -\frac13 m_\mu \left(1+O\left(\frac{f_3}{h_3}\right)\right).
\end{eqnarray}
Since the relations are satisfied
under the assumption of approximate rank 1 property
irrespective of the counting of freedom,
they are stable
even in the case of three generation model.
%or when we add $h^\prime$ as a small correction.
Indeed, the predictivity from the minimality of the parameter
is related to the $O(f_3/h_3)$ corrections,
and the minimality does not play a crucial role in the approximate relations
from the rank 1 assumption.

It is known that there is a solution that the large atmospheric mixing
is obtained even if the smallness of $f$ is not assumed a priori.
In the scenario, the $b$-$\tau$ mass convergence as well as the other
experimental
inputs predict the neutrino mixing as an output \cite{goran}.
Our main goal in this section is not to give numerical predictions but
rather to
show how one can get qualitatively expected hierarchical pattern for
masses and mixings. Later
on we of course study the detailed numerical predictions.
As it turns out there is a fine-tuned solution to fit the experimental
data even if $f_3$ is comparable to $h_3$. however such
a fine-tuned solution is sensitive to the numerical inputs,
and therefore the numerical predictions in this case may be unstable
under a possible higher order correction.
%
%In fact, in the models with ${\bf 126}+\overline{\bf 126}$ Higgs fields,
%the gauge couplings may blow up just above the unification scale,
%and thus the higher order corrections should be taken into account.
%In the sense, the stability of the solution is important, rather than
%the minimality of the parameters.
%
%We discuss the stable relations among fermion masses and mixings
In our case where the quark and lepton mass hierarchy is predicted
by the rank 1 assumption, they are stable under radiative corrections.

While the qualitative predictions are in the expected range, we note that
the approximate relation $\tan \theta_{\rm atm} \simeq
(m_s/m_b)/V_{cb}$
is not very good agreement with the current observation,
and small $h^\prime$ will be invoked to obtain the best fit of the
experimental data.
%
%However, what we want to emphasize here is not the numerical predictivity originating from
%the minimality of the number of parameters.
%We have illustrated the consequence of the rank 1 property of the charged fermion masses.
We emphasize that our final solutions do not use any fine tuned
cancellations, and thus
are stable even if we add small corrections to fit the numerical
experimental data.
%

%This model should have three predictions by the afore mentioned
%parameter counting. They are
%\begin{eqnarray}
%\frac{m_s}{m_b}~\simeq -V_{cb} tan\theta\\ \nonumber
%m_\tau~\simeq m_b\\ \nonumber
%m_\mu~\simeq -3 m_b
%\end{eqnarray}
%All these predictions are good agreement with observations
%indicating that the ansatz is taking us in the right track for
%understanding quark-lepton flavor. We will see below that a
%similar situation happens for the three generation case.

\subsection{Three generation case}

The fermion mass equations for this case are those in Eq. (\ref{eq2})
with all coupling matrices being $3\times 3$. The assumption
that $h$ has rank one means that we can write it as
\begin{equation}
h = \left(
       \begin{array}{c} c \\ b \\ a
       \end{array}
      \right)
      \left(
       \begin{array}{ccc} c & b & a
       \end{array}
      \right),
\end{equation}
\begin{equation}
f = {\rm diag} (f_1 ,f_2, f_3)
\qquad (f_{1,2} \ll f_3).
\end{equation}
Again, we can parameterize $f$ to be diagonal and $a,b,c$ to be real
without loss of generality.
At first, we ignore $h^\prime$.
In order to analyze the detailed consequences of this assumption, we go to
the basis where $h$ is diagonal. This is achieved by the matrix:
\begin{equation}
U_0 = \left( \begin{array}{ccc}
                 \cos \theta_s &  \sin\theta_s & 0 \\
                 -\cos \theta_a \sin\theta_s & \cos\theta_a
\cos\theta_s & -\sin\theta_a \\
                 -\sin \theta_a \sin\theta_s & \sin\theta_a
\cos\theta_s & \cos\theta_a
               \end{array}
        \right),
\label{U0}
\end{equation}
where $\tan\theta_s = -c/b$ and $\tan\theta_a =
\sqrt{b^2+c^2}/a$
with
\begin{eqnarray}
U_0 h U_0^{t} &=& {\rm diag} (0,0,
h_3),
\end{eqnarray}
where $h_3 = a^2+b^2+c^2$.
It is interesting to note
that in the diagonalization matrix there is an ambiguity
resulting from the residual SU(2) flavor symmetry in $h$
(i.e. one of three mixing angles is not fixed at this stage).
We choose the unitary matrix $U_0$ to be an approximate leading order
diagonalization matrix of $Y_u$, $Y_d$, and $Y_e$ as in the previous
subsection
($V_u = \tilde V_u U_0$, $V_d = \tilde V_d U_0$, and $V_\ell = \tilde
V_e U_0$
where $\tilde V_u,\tilde V_d,\tilde V_e$ are close to a unit matrix).
Then, once the $f$ contribution is included, the afore mentioned SU(2)
flavor symmetry is broken
and the ambiguity in mixing angles alluded to above is removed. We wish
to note that in our original parameterization of $U_0$, we chose the 13
element to be zero since even after including the $f$-contribution,
 the 13 element goes to zero in the limit $f_{1,2}/f_3 \to
0$ (which is the limit where $Y_{u,d,e}$ is rank 2).
We have not used prejudices from neutrino experiment.

%We have assumed that the matrix $h$ is real and symmetrical.
By the same argument as in the case of two generations,
$U_0$ is cancelled out in the CKM mixing matrix
and the quark mixings are small.
The PMNS matrix is given by
\begin{eqnarray}
U_{\rm PMNS}~=~\tilde{V}_e^* U_0,
\end{eqnarray}
 and since the off-diagonal
elements of $\tilde{V}_\ell$ are small (being related to quark mixings),
neglecting the $23$ and $13$ quark mixings, we get for the solar and
atmospheric mixing angles \cite{dutta}
\begin{eqnarray}
&&\theta_{\rm atm}\simeq \theta_a, \label{th-sol} \\ \nonumber
&&\theta_\odot\simeq \theta_s\pm \theta_{13}\cot\theta_a \cos\alpha,
\end{eqnarray}
where $\alpha$ is defined as the diagonal phase matrix ${\rm diag}(1,
e^{i\alpha}, e^{i\beta})$ needed to diagonalize the charged lepton mass
matrix \cite{footnote}.

We also get a formula for $U_{e3}$ as follows:
\begin{eqnarray}
U_{e3}~=~(\tilde{V}_e)_{12} \sin\theta_a.
\label{ue3-v12}
\end{eqnarray}

To proceed with the rest of the masses and mixings,
let us define the matrices in the $U_0$ rotation:
$\tilde Y_a \equiv U_0 Y_a U_0^t$,
$\tilde f \equiv U_0 f U_0^t$, and so on.
In this notation, $\tilde V_a$ is a diagonalization matrix of $\tilde Y_a$.
Because $\tilde f_{23} = (f_2- f_3) \sin\theta_a \cos\theta_a$
and $\tilde f_{13} = (f_2- f_1) \sin\theta_a \sin\theta_s \cos\theta_s$,
one can obtain
\begin{equation}
V_{ub} \simeq V_{cb} \frac{f_2-f_1}{f_3} \frac{\sin\theta_s
\cos\theta_s}{\cos\theta_a}.
\end{equation}

Neglecting $O(f_3/h_3)$ and $O(f_{1,2}/h_3)$ corrections,
$\tilde V_a$ can be approximately
as:
\begin{equation}
\check V = \left(
 \begin{array}{ccc}
   \cos\check\theta & -\sin\check\theta & 0 \\
   \sin\check\theta & \cos\check\theta & 0 \\
   0 & 0 & 1
 \end{array}
\right),
\label{check-V}
\end{equation}
where
\begin{equation}
\sin\check\theta \simeq \frac{f_2-f_1}{f_3} \frac{\cos\theta_a
\sin\theta_s\cos\theta_s}{\sin^2\theta_a},
\label{check-th}
\end{equation}
and thus
\begin{eqnarray}
U_{e3}~\simeq~\frac{f_2-f_1}{f_3}\cot\theta_a \sin\theta_s\cos\theta_s.
\end{eqnarray}
Thus we have obtained all the features
listed before.
%As we have emphasized, the crucial assumption is rank 1 property
%and numerical inputs do not play a key role to obtain the approximate features.
%
Due to the generic largeness of the relative mixing angles of the
unrelated matrices,
solar and atmospheric neutrino mixing angles are of $O(1)$ generically.
On the other hand, 13 mixing is not in the category of the generic
largeness since it is
related to the ratio of eigenvalues of $f$.
The eigenvalue ratio is also related to $V_{ub}/V_{cb}$ implying that
the 13 mixing angle has to be small in our approach.
It is important to note
that we do not assume a particular flavor texture such like hierarchical
pattern in one matrix to obtain the feature.
The key property to obtain the features for the neutrino mixings is
that the correction to the rank one charged lepton mass matrix
and the type II seesaw term are unified (or more roughly, simultaneously
diagonalized), as a result of SO(10) unification.

\subsection{Realistic model with $h^\prime$}

The discussion above gives the qualitative consequences of the rank one
property, and the experimental inputs are not used to obtain the
features.
The discussion below will address the issue of the experimental
data
for the first generation.
Actually, we have not listed the first generation masses and $V_{us}$ before.
In fact,
if $h^\prime =0$,
one obtains the following relation among the fermion masses:
\begin{eqnarray}
 \frac{m_u}{m_t} : \frac{m_d}{m_b} : \frac{m_e}{m_\tau}
\simeq r_2 (1+r_2 X): 1+X: -3(1-3X),
\end{eqnarray}
where $X= f_1f_2f_3/(a^2 f_1 f_2+ b^2 f_1 f_3+ c^2 f_2 f_3)$.
When one fits down quark and electron masses (e.g., $X = 0.35$), the
up quark mass is clearly too large since $r_2 \simeq m_c/m_t/(m_s/m_b)
\sim 0.1$.
As a result, one of the first generation masses cannot be fitted.
Besides, since $\check V$ in Eq.(\ref{check-V}) is common for up- and
down-type quarks,
$V_{us}$ becomes too small compared to observations, since $V_{us} \simeq
V_{cb} V_{ub}$.
Therefore, one needs non-vanishing contribution from $h^\prime$ to obtain
realistic masses for the first generation and $V_{us}$ under the rank 1
assumption.

As is well-known,
the empirical relation $V_{us} \simeq \sqrt{m_d/m_s}$
is obtained when $(\tilde Y_d)_{11} \to 0$ and $(\tilde Y_d)_{12} \simeq
(\tilde Y_d)_{21}$.
Therefore we choose $\tilde f_{11} \to 0$.
When $(\tilde Y_d)_{11}, (\tilde Y_e)_{11} \to 0$ is assumed,
the choice of $(\tilde Y_e)_{12} (\tilde Y_e)_{21} \sim (\tilde
Y_d)_{12} (\tilde Y_d)_{21}$
satisfies the Georgi-Jarskog (GJ) relation ($m_e m_\mu m_\tau \sim m_d
m_s m_b$)
for the down-type quarks and charged lepton masses.
The up quark mass can be fit by using the freedom of $r_3$.
As a result,
we have the following two solutions typically to fit the first
generation masses
and $V_{us}$ in a simple manner.

\bigskip

\noindent {\it Case A}:
$\tilde f_{11} \simeq 0$ and
$|\tilde f_{12} + \tilde h^\prime_{12}|\simeq|-3\tilde f_{12} + \tilde
h^\prime_{12}|$.
The smallness of up quark mass is
realized by a cancellation in $(\tilde Y_u)_{12} = r_2 \tilde f_{12} +
r_3 \tilde h^\prime_{12}$.

In this case, $h^\prime$ has to be symmetric which we can obtain by
employing an extra ${\bf 10}$ Higgs field.
For example, $\tilde h^\prime_{12} \simeq \tilde f_{12}$
is the simplest solution, giving
\begin{eqnarray}
U_{e3} &\simeq& \frac13 V_{us} \sin \theta_{a}, \\
V_{us} &\simeq& 2 \sin \check\theta
\simeq 2 \frac{f_2}{f_3} \frac{\cos\theta_a}{\sin^2\theta_a} \tan\theta_s.
\end{eqnarray}
where $\check\theta$ is given in Eq.(\ref{check-th})
and used a relation $f_1 \simeq -f_2 \tan^2\theta_s$ from $\tilde f_{11}
\simeq 0$.
Assuming that the corrections from the other elements of $\tilde h^\prime$
(e.g., $\tilde h^\prime_{13,23}$)
are small,
we have an approximate relation
\begin{equation}
\frac{V_{ub}}{V_{cb}} \sim \frac12 V_{us} \tan^2\theta_{a},
\end{equation}
which is in good agreement with the experiment.

\bigskip

\noindent{\it Case B}:
In this case we have $\tilde f_{11}\tilde f_{22}-\tilde f^2_{12}~\simeq~0$.
%In this case, we consider a situation where after a rotation by $\check
%V$, the 12 element of $\check V \tilde f
%\check V^t$ is zero whereas the
%So if we choose $\tilde f_{11}=0$, then this implies that $\tilde
%f_{12}=0$.
Then,
11 and 12 elements of $(\check V \tilde f \check V^t)$ are zero,
where $\check V$ in Eq.(\ref{check-V}) is an approximate diagonalization
matrix in the limit $h^\prime \to 0$.
The 12 element of $\check V \tilde h^\prime \check V^t$ produces the Cabibbo angle.
The GJ relation is manifest when $|c_e|=1$
and $(\check V \tilde h^\prime \check V^t)_{11} \simeq 0$.
The up quark mass is fitted by the smallness of $r_3$,
$m_u/m_c \simeq r_3^2/r_2 m_d/m_s$.

In this case, $h^\prime$ can be either symmetric or anti-symmetric.
Since 11 element vanishes automatically, anti-symmetric coupling from {\bf 120}
Higgs field is a better choice.
As is noted, $\check V$ contributes to $U_{e3}$,
but it does not contribute to the Cabibbo angle.
As a result, we obtain from Eq.(\ref{ue3-v12})
\begin{equation}
|U_{e3}| \simeq \sin\theta_a |\sin\check \theta + e^{i\gamma} c_e
\frac13 V_{us}|
\simeq
\left|\frac{f_2}{f_3} \tan\theta_s \cot\theta_a + e^{i \gamma} c_e
\frac13 V_{us} \sin\theta_a
\right|,
\label{Ue3-case-B}
\end{equation}
where $\gamma$ is a relative phase between $\tilde f_{12}$ and $\tilde
h_{12}^\prime$ roughly,
and we have used a relation $f_1 \simeq -f_2 \tan^2\theta_s$.
Since $V_{us}$ is generated purely from $h^\prime$,
it is not directly correlated to $V_{ub}/V_{cb}$ contrary to the case A.

%Note that if $h^\prime=0$, one obtains the following relation among
%fermion masses:
% \begin{eqnarray}
% \frac{m_u}{m_t} : \frac{m_d}{m_b} : \frac{m_e}{m_\tau}
%\simeq r_2 : 1: -3
%\end{eqnarray}
%which clearly implies that the electron mass is too large. The case with anti-symmetric
%$h^\prime$ is
%in fact the model discussed in \cite{mimura} which as was shown there is not
%only realistic but it also predicts $U_{e3}\geq 0.05$ and also a large Dirac phase.

\bigskip

As we have noted, to fit $V_{cb}$, $m_s/m_b$ and $\theta_{\rm atm}$ very well,
one needs a correction in $\tilde h^\prime_{23}$.
However, the correction does not affect the approximate expressions for
$U_{e3}$ very much.

It is interesting that the $U_{e3}$ is related to the mass ratio
of neutrino in both cases.
Since the GJ relation and the empirical relation of $V_{us}$ are not
exact relations,
there can be a shift from them in a numerical fit analysis.

Here we assumed $(\tilde Y_a)_{11} \to 0$ to
satisfy the GJ relation and the empirical relation of $V_{us}$
in a simple manner.
When one introduces other parameters especially for symmetric $h^\prime$,
there will be an accidental fine-tuned solution
for the relations in a general fit for $(\tilde Y_a)_{11}\neq 0$.
Actually, when $h^\prime$ is symmetric ($c_e=1$) and $r_3=1$,
it results in a minimal model in which only one ${\bf 10}$
and $\overline{\bf 126}$ Higgs fields couple to fermions
with $h$ being rank 3.
In the minimal model for the fermion sector,
it is known that there is a fine-tuned solution
to fit fermion masses and mixings \cite{Babu:2005ia}
unless the minimality of the Higgs potential is taken into account.
In this case, when the first generation masses and $V_{us}$ are
tuned,
there is no freedom to adjust $U_{e3}$
and thus the approximate relation in the previous subsection holds,
$U_{e3} \sim f_2/f_3$.
When $r_3 \neq 1$ (but $r_3 \simeq 1$),
$U_{e3}$ can be tuned to be any value (including zero)
since first generation masses and $V_{us}$ can be fitted
even if $(\tilde Y_e)_{12}=0$.
(When $h$ is rank 1 and $h^\prime$ is anti-symmetric,
there is no such fine-tuned solution.
When $h$ is rank 3, the fine-tuning fit for $U_{e3}=0$ is allowed
\cite{Dutta:2004wv}.)
Therefore, the assumption $(\tilde Y_a)_{11} \to 0$
to satisfy GJ relation in a simple manner
is crucial to keep the $U_{e3}$ prediction.
Actually, when $(\tilde Y_a)_{11} \to 0$ is assumed,
the fine-tune solutions are removed, and the $U_{e3}$ is predicted
as we have noted, irrespective of the number of parameters.
We also note that the
assumption $(\tilde Y_a)_{11} \to 0$
is preferable to suppress nucleon decay amplitudes
naturally.

The case with the assumption
that $h$ (rank 1) and $f$ are real and anti-symmetric $h^\prime$ is pure
imaginary
(in which case, the charged fermion mass matrices are hermitian)
is in fact the model discussed in \cite{mimura}.
For this case, cancellation cannot happen between $f$ and $h^\prime$
and thus the numerical fit does not shift very much from the above expression.
In the numerical fit, it predicts $|U_{e3}| = 0.08- 0.12$ in the case
where $|c_e| =1$
(and the GJ relation is manifest).
Under the hermiticity assumption, one obtains $e^{i\gamma} = \pm i$
and it is consistent with the above expression.
In this case, since $(\tilde Y_e)_{22}$ is real,
the PMNS phase is roughly same as the phase in the expression in
Eq.(\ref{Ue3-case-B})
and thus
\begin{equation}
\tan \delta_{\rm PMNS} \simeq \frac13 \frac{c_e V_{us}}{\sin\check\theta},
\end{equation}
and we obtain $\delta_{\rm PMNS} \simeq \pm 30^{\rm o}$ or $180 \pm
30^{\rm o}$
using the experimental inputs.

For the case of $|c_e|\neq 1$, one can also fit the experimental data
and the prediction is $|U_{e3}| = 0.05- 0.14$.
Since the cancellation is not allowed between $\tilde f_{13}$
and $\tilde h^\prime_{13}$,
the experimental data of $V_{ub}$ cuts the upper region
of experimentally allowed mass squared ratio difference,
and then gives an upper bound of $U_{e3}$.

\section{Tri-bimaximal ansatz}
In the previous section, we incorporated large lepton mixings but their
values were inputs into the theory. In this section, we consider special
cases where the dominant part of the lepton mixing is in the
tri-bimaximal form \cite{tbm}. This would require special form for the rank one
matrix $h$. We envisage that the rank one form for $h$ as well as the
matrix forms
for $f$ and $h'$ come from some vacuum alignment of
flavon fields, e.g., \cite{Ross,Luhn:2007sy}.

In the triplet flavon models,
the $3\times 3$ matrix can be expanded by the tensor products
of the flavon fields
when there are three independent flavons.
The three flavon fields
can be expressed (without loss of generality, by making unitary
transformations) as
\begin{equation}
\phi_1 = (0,0,1), \quad \phi_2 = (0,a,b), \quad \phi_3 =
(c,d,e).
\end{equation}
In general, there is no reason for the flavon vevs to be
hierarchical,
and the large neutrino mixings can originate from $a\sim b$, $c\sim d
\sim e$ \cite{Kitano:2000xk}.
The experimental result from the neutrino oscillation
seems to imply a special alignment of flavon vevs
rather than the generic largeness,
namely \cite{Ross},
\begin{equation}
\phi_1 = (0,0,1), \quad \phi_2 = (0,-1,1)/\sqrt2, \quad \phi_3 =
(1,1,1)/\sqrt3.
\label{alignment}
\end{equation}
The vacuum alignment can be obtained by imposing a discrete flavor
symmetry \cite{Luhn:2007sy}, leading to tri-bimaximal neutrino mixings.

%
%It is worth pointing out at the beginning that in the rank one case, the Yukawa
%coupling can be written in several ways by choice of basis (or by making unitary transformations).
% The final result is independent of the basis choice.
%We start with the form of the flavon fields in terms of which we will
%give our Yukawa couplings: The first basis we mention is:
%\begin{equation}
%\phi_1 = (0,0,1), \quad \phi_2 = (0,-1,1)/\sqrt2, \quad \phi_3 =
%(1,1,1)/\sqrt3.
%\end{equation}
%

It is worth pointing out at the beginning that
the aligned flavon fields can be written in several ways by choice of
coordinates (or by making unitary transformations) and the final results
are independent of the coordinate choice.

It is interesting to note that the aligned flavon vevs
correspond to a link of a hexahedron, a diagonal line of a lateral
surface, a diagonal line of a regular hexahedron,
respectively.
The interpretation becomes clear when the flavon fields
are expanded in terms of the following orthogonal axes of coordinates
(called
hexahedral coordinate)
\begin{equation}
x_1 = (1,0,0),\quad x_2 = (0,1,0), \quad x_3 = (0,0,1),
\end{equation}
which correspond to the three lateral links of the regular hexahedron.
The hexahedral coordinate is convenient to describe the $Z_4$ rotation
around the surface-diagonal axes of the hexahedron.
In fact, the regular hexahedron has different coordinates
to describe the symmetry of the shape.
One can consider a coordinate system which proves convenient
to describe the $Z_3$ rotation around the vertex-diagonal axes
of the hexahedron,
%
%
%The regular hexahedron has different coordinates
%to describe the symmetry of the shape. %where the diagonal lines are two
%%of the axes.
%%
%The coordinate for the lateral links of the hexahedron (called hexahedral
%basis)
%\begin{equation}
%x_1 = (1,0,0),\quad x_2 = (0,1,0), \quad x_3 = (0,0,1).
%\end{equation}
%
%
\begin{equation}
x_1^\prime = (2,-1,-1)/\sqrt6,\quad x_2^\prime = (1,1,1)/\sqrt3, \quad
x_3^\prime = (0,-1,1)/\sqrt2.
\end{equation}
%
%The coordinate for the diagonal lines of the hexahedron (
The axes of the coordinates $x_1^\prime$ and $x_3^\prime$
are on the regular triangle which is formed by three of the hexahedron's
vertices, and $x_2^\prime$
is perpendicular to the triangle.
%\begin{equation}
%x_1^\prime = (2,-1,-1)/\sqrt6,\quad x_2^\prime = (1,1,1)/\sqrt3, \quad
%x_3^\prime = (0,-1,1)/\sqrt2.
%\end{equation}
%
We call this tetrahedral coordinate.

The unitary matrix for the coordinate transformation from hexahedral
(unprimed) to tetrahedral (primed)
coordinate is the
tri-bimaximal (TB) matrix i.e. $x'=x U^t_{\rm TB}$, where
\begin{equation}
U_{\rm TB} = \left( \begin{array}{ccc}
                 \sqrt{\frac23} & \sqrt{\frac13} & 0 \\
                 -\sqrt{\frac16}  & \sqrt{\frac13}& -\sqrt{\frac12} \\
                 -\sqrt{\frac16} & \sqrt{\frac13} & \sqrt{\frac12}
               \end{array}
        \right).
\end{equation}
Therefore, in general (irrespective of the rank one assumption), if
the charged-lepton mass matrix is (nearly) diagonal in the hexahedral
coordinate
and the neutrino mass matrix is (nearly) diagonal in the tetrahedral
coordinate,
then the neutrino mixing matrix is (nearly) tri-bimaximal and given by:
\begin{equation}
U_{\rm MNSP} = V_e U_{\rm TB} V^\dagger_\nu,
\end{equation}
where $V_e$ is a diagonalizing matrix of $Y_e$
in the hexahedral coordinate,
and $V_\nu$ is a diagonalizing matrix of ${\cal M}_\nu$
in the tetrahedral coordinate.

%Suppose we assume that the vacuum alignment in the tetrahedral coordinate
%is given by

Suppose that the vacuum alignment in Eq.(\ref{alignment}) is given in the
hexahedral coordinate as we have noted.
Then, those flavons in the tetrahedral coordinate are given as ($\phi_i^\prime = \phi_i U_{\rm TB}$)
\begin{equation}
\phi_1^\prime = (-1,\sqrt2,\sqrt3)/\sqrt6, \quad \phi_2^\prime = (0,0,1),
\quad \phi_3^\prime = (0,1,0).
\end{equation}
Therefore,
from the discussion in the previous section, one can easily check
that
the nearly tri-bimaximal neutrino mixings are obtained
when $h$ is rank one formed by $\phi_1$ (irrespective of the choice of
the coordinate),
and $f$ is formed by $\phi_2$ and $\phi_3$.
We define $\phi_4$ which is obtained an outer product of $\phi_2$ and $\phi_3$,
i.e. $\phi_4 = \phi_3 \times \phi_2$.
In the tetrahedral coordinate, $\phi_4^\prime = (1,0,0)$.

As we have mentioned,
the Yukawa matrices can be expressed in terms of the tensor products of
the flavon fields.
The symmetric matrices can be formed by six bases.
Since we set 11 element to be zero in the hexahedral coordinate,
we define the following matrices as the bases to form the linear space
of symmetric matrices:
%
%In order to present the two cases that are of interest to us, we start
%by defining some basis matrices:
\begin{eqnarray}
Y_1 &=& \phi_1^t\phi_1 =
\left(
\begin{array}{ccc}
0 & 0 & 0 \\
0 & 0 & 0 \\
0 & 0 & 1
\end{array}
\right), \\
Y_2 &=& 2 \phi_2^t \phi_2 =
\left(
\begin{array}{ccc}
0 & 0 & 0 \\
0 & 1 & -1 \\
0 & -1 & 1
\end{array}
\right), \\
Y_3 &=& 2 (\phi_3^t \phi_3-\frac12 \phi_4^t \phi_4) =
\left(
\begin{array}{ccc}
0 & 1 & 1 \\
1 & \frac12 & \frac12 \\
1 & \frac12 & \frac12
\end{array}
\right), \\
Y_4 &=& \sqrt6 (\phi_2^t \phi_3 + \phi_3^t \phi_2) =
\left(
\begin{array}{ccc}
0 & -1 & 1 \\
-1 & -2 & 0 \\
1 & 0 & 2
\end{array}
\right), \\
Y_5 &=& \sqrt3 (\phi_2^t \phi_4 + \phi_4^t \phi_2) =
\left(
\begin{array}{ccc}
0 & -1 & 1 \\
-1 & 1 & 0 \\
1 & 0 & -1
\end{array}
\right),
\end{eqnarray}
where the elements of the matrices are presented in the hexahedral
coordinate.

In the following, we will consider two models:
 (I) $V_\nu$ is a unit matrix ($f$ is diagonal in the tetrahedral
coordinate),
(II) $V_\nu$ is close to a unit matrix.

\subsection{Model I :  $V_\nu = {\bf 1}$}
%This corresponds to the choice:
The Model I can have both case A and case B solutions as described in
the previous section. The case A solutions are,
however, more natural in this model.
In order to obtain such a  solution,
we employ additional ${\bf 10}$ Higgs to obtain a correction matrix
$h^\prime$,
and $h^\prime$ is a symmetric matrix.

We arrange the $h$, $f$, $h^\prime$ couplings as follows:
\begin{eqnarray}
h &=& h_3\, Y_1, \\
f &=& h_3\, \epsilon\,(Y_2 + \lambda\, Y_3), \\
h^\prime &=& h_3\, \epsilon\, \lambda\, \rho\, Y_4 \quad ({\rm or} \
h_3\, \epsilon\, \lambda\, \rho\, Y_5).
\end{eqnarray}

Then, since the ratio of eigenvalues of $f$ is $1:\lambda:-\lambda/2$,
we obtain $\Delta m^2_{\rm sol}/\Delta m^2_{\rm atm} = \frac34 \lambda^2$.
In the parameterization in the previous section,
$U_0$ in Eq.(\ref{U0}) is the tri-bimaximal matrix because
$a:b:c = \sqrt3:\sqrt2:-1$.

The fermion Yukawa matrices are
\begin{eqnarray}
Y_u &=& h+r_2 f + r_3 h^\prime \\
&=&
h_3 \left(
\begin{array}{ccc}
0 & \epsilon\lambda (r_2 - r_3 \rho) & \epsilon\lambda (r_2 + r_3 \rho) \\
\epsilon\lambda (r_2 - r_3 \rho) & r_2 \epsilon(1+ \frac{\lambda}2) +
r_3 x \epsilon\lambda\rho &
-\epsilon r_2 (1-\frac{\lambda}2) \\
\epsilon\lambda(r_2 + r_3 \rho) & -\epsilon r_2 (1-\frac{\lambda}2) & 1+
r_2 \epsilon (1+\frac{\lambda}2)
 -r_3 x \epsilon\lambda\rho
\end{array}
\right), \\
Y_d &=& r_1(h + f + h^\prime) \\
&=&
r_1 h_3 \left(
\begin{array}{ccc}
0 & \epsilon\lambda (1 - \rho) & \epsilon\lambda (1 +  \rho) \\
\epsilon\lambda (1 - \rho) & \epsilon(1+ \frac{\lambda}2) + x
\epsilon\lambda\rho &
-\epsilon  (1-\frac{\lambda}2) \\
\epsilon\lambda(1 + \rho) & -\epsilon (1-\frac{\lambda}2) & 1+ \epsilon
(1+\frac{\lambda}2)
 -x \epsilon\lambda\rho
\end{array}
\right), \\
Y_e &=& r_1(h -3 f + h^\prime) \\
&=&
r_1 h_3 \left(
\begin{array}{ccc}
0 & \epsilon\lambda (-3 - \rho) & \epsilon\lambda (-3 + \rho) \\
\epsilon\lambda (-3 - \rho) & -3 \epsilon(1+ \frac{\lambda}2) + x
\epsilon\lambda\rho &
3\epsilon (1-\frac{\lambda}2) \\
\epsilon\lambda(-3 + \rho) & 3\epsilon (1-\frac{\lambda}2) & 1-3
\epsilon (1+\frac{\lambda}2)
 -x \epsilon\lambda\rho
\end{array}
\right),
\end{eqnarray}
where $x = -2$ when $h^\prime \propto Y_4$,
and $x = 1$ when $h^\prime \propto Y_5$.

For the numerical fits,
the parameter $\epsilon$ is given by  $\epsilon \sim m_s/m_b \sim V_{cb}$,
and the parameter $\lambda$ is given by $\lambda(1-\rho) \sim V_{us}$.

When $\rho \simeq -1$, then
Goergi-Jarskog relation is satisfied naturally.
At that time, $(Y_d)_{13} \simeq 0$.
This is interesting since
the empirical relation
$|V_{td}| \simeq V_{us} V_{cb}$ is satisfied simultaneously
when $r_2, r_3$ are small (up-type quark masses are more hierarchical
rather than down-type quark masses).

The parameter $r_2$ is fixed as $|r_2| \sim m_c/m_t/(m_s/m_b)$.
The up quark mass can be made small
by a choice $r_3 \sim r_2 /\rho$.

Cabibbo angle, $U_{e3}$ and the ratio of mass squared differences
$\Delta m^2_{\rm sol}/\Delta m^2_{\rm atm}$
are all correlated by the parameter $\lambda$.
The naive approximate relation is $U_{e3} \simeq  V_{us}/(3\sqrt2)$
as we have derived in the previous section.
%According to our numerical calculation,
%$U_{e3} \simeq 0.06-0.07$.
%A smaller side of experimental value of the ratio $\Delta m^2_{\rm
%sol}/\Delta m^2_{\rm atm}$
%(larger value of $\Delta m^2_{\rm atm}$) is favored,
%which can be understood by
%the native relation $V_{us} \sim 2 \lambda$,
%and $\Delta m^2_{\rm sol}/\Delta m^2_{\rm atm}= 3\lambda^2/4$

%In general, all the coefficients are complex.
%But the combination of $ab$ is real without loss of generality
%in the first row and column.
%For example, if we take the Georgi-Jarskog relation and
%$V_{td} \simeq V_{us} V_{cb}$ into account,
%$c \simeq -1$ and $c$ is real.
%Then,
%the KM phase comes from $b$ or $r_2$ (or those combination).
%If
%only $r_2$ contribute the KM phase and $b$ is real,
%the MNSP phase for neutrino oscillation should be 0 or $\pi$.
%If the Higgs mixing $r_2$ is real and $b$ is complex,
%the KM phase and MNSP phase are correlated.
\begin{figure}
\centering
\mbox{\subfigure{\includegraphics[viewport = 10 15 280 220,width=3in]{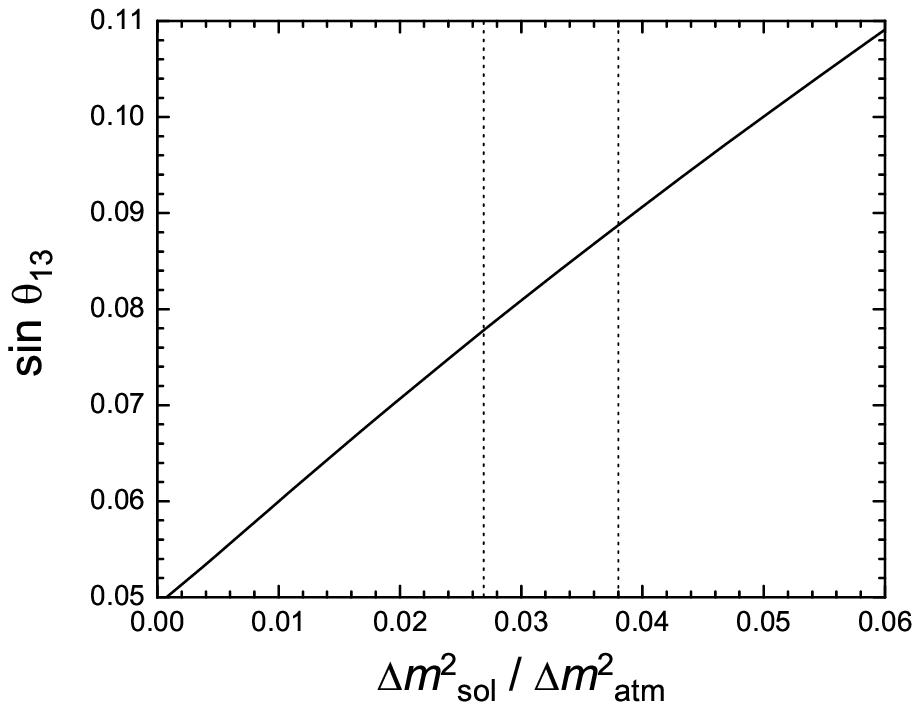}}\quad
\subfigure{\includegraphics[viewport = 10 15 280 220,width=3in]{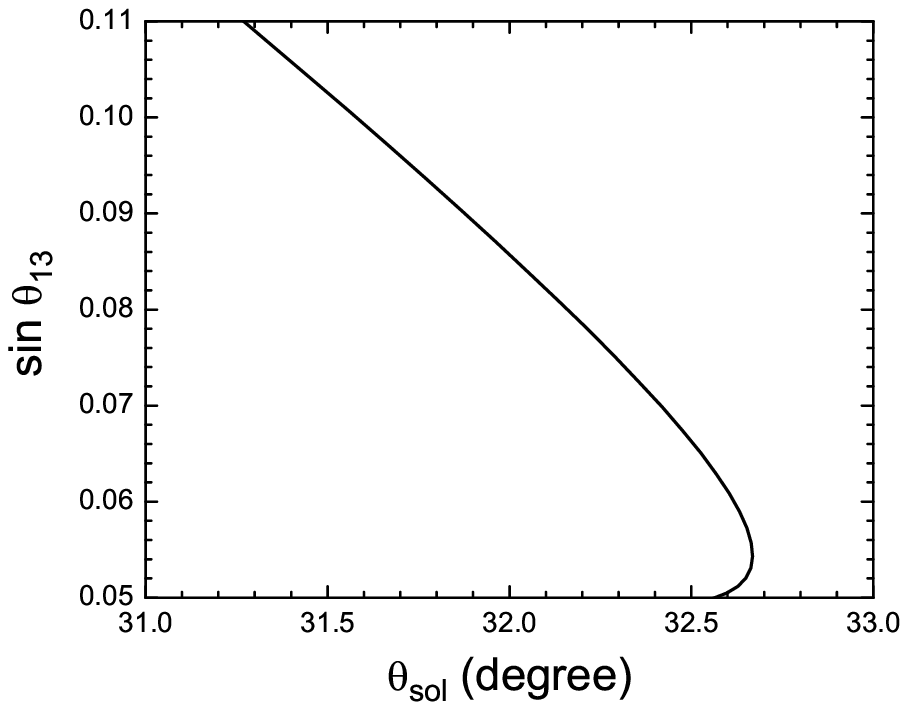} }}
\caption{$U_{e3}$ is shown as a function of $\Delta m^2_{\rm
sol}/\Delta m^2_{\rm atm}$ (left) and $\theta_s$(right) for Model I.}
\label{fig12}
\end{figure}

In Fig. 1, we plot $U_{e3}$ as a function of the mass squared difference
ratio.
In the plot, we fit $m_e/m_\tau$ and $m_\mu/m_\tau$ using $\rho$ and
$\epsilon$
(which are assumed to be real in the plot).
Then, $U_{e3}$ is calculated as a function of $\lambda$.
(The mixing angles do not depend on $h_3$).
We note that
a correction from $h^\prime$ is needed
to fit $V_{cb}$, e.g., $\Delta h^\prime \propto \phi_1^t \phi_2 +
\phi_2^t\phi_1$.
As we have mentioned, such correction do not modify the $U_{e3}$ very much.
In Fig. 1, we also show the plot of $U_{e3}$ as a function of
$\theta_{\rm sol}$.
Using the experimental constraint on $\Delta m^2_{\rm sol}/\Delta
m^2_{\rm atm}$
and the other input from quark masses and mixings,
we find that $U_{e3}$ is predicted to be 0.07-0.08.
We note that the smaller side of experimental range of $\Delta m^2_{\rm
sol}/\Delta m^2_{\rm atm}$
is preferred from the numerical fit,
which obeys from the naive relation $\lambda (1-\rho)\sim V_{us}$.
The solar mixing angle $\theta_{\rm sol}$
is found to be $\sim 32^{\rm o}$ from the plot
which is obeyed by the approximation Eq.(\ref{th-sol})
when the PMNS phase is 0 (or $\pi)$, which is resulting from the
assumption where
$\rho$ and $\epsilon$ are real.
The atmospheric mixing angle $\theta_{\rm atm}$ is $45^{\rm o}$ up to
$\pm$2-3$^{\rm o}$ correction
from $V_{cb}$
for Model I irrespective of case A or case B solutions.

The current allowed range for the neutrino parameters at 2$\sigma$ level
are as follows~\cite{Fogli:2008cx}:
$\theta_{\rm atm}= 37^{\rm o}-51^{\rm o}$, $\theta_{\rm sol}=31.8^{\rm
o}-36.4^{\rm o}$ and
$\Delta m^2_{\rm sol}/\Delta m^2_{\rm atm}=0.027-0.038$.

\subsection{Model II: $V_\nu \neq 1$ }

The Model II can have both case A and case B solutions as well. In this
model, the case B solutions
are more natural.
It is possible that the $f$ coupling is not completely diagonal in the
tetrahedral coordinate.
Using the available freedom, we choose the 12 element of $f$ in the
hexahedral coordinate
to obtain the case B solution.
Since the 12 elements of $Y_3+Y_4$ and $Y_3+Y_5$ are zero,
%
%One can consider the case $f$ has a correction.
%As is mentioned in the previous section,
%let us consider the corrections to make $f_{12}$ is zero
%in the unprimed basis.
%
one can consider the choice:
\begin{eqnarray}
h &=& h_3\, Y_1, \\
f &=& h_3\, \epsilon\,\left(Y_2 + \lambda\, (Y_3+ Y_4)\right), \\
h^\prime &=& h_3\, \epsilon\, \lambda\, \rho \, Y_4.
\end{eqnarray}
The matrix $Y_4$ can be replaced with $Y_5$.
One can also choose $h^\prime$ to be antisymmetric,
e.g., $h^\prime \propto \phi_2^t\phi_3-\phi_3^t\phi_2$.

%\begin{eqnarray}
%Y_u &=& h+r_2 f + r_3 h^\prime \\
%&=&
%y \left(
%\begin{array}{ccc}
%0 & -abc^\prime r_3 & ab (2r_2 + r_3 c^\prime) \\
%-abc^\prime r_3  & r_2 b(1+ \frac{a}2+xa) + r_3 x a b c^\prime & -b r_2
%(1- \frac{a}2) \\
%ab(2r_2 + r_3 c^\prime) & -b r_2 (1-\frac{a}2) & 1+ r_2 b
%(1+\frac{a}2-xa) -r_3 x abc^\prime
%\end{array}
%\right), \\
%%
%Y_d &=& r_1(h + f + h^\prime) \\
%&=&
%r_1 y \left(
%\begin{array}{ccc}
%0 & -ab c^\prime & ab (2+ c^\prime) \\
%-ab c^\prime & b(1+ \frac{a}2+xa) + x a b c & -b (1- \frac{a}2) \\
%ab(2 + c^\prime) & -b (1-\frac{a}2) & 1+ b (1+\frac{a}2-xa) -x abc
%\end{array}
%\right), \\
%%
%Y_e &=& r_1(h -3 f + h^\prime) \\
%&=&
%r_1 y \left(
%\begin{array}{ccc}
%0 & -abc^\prime & ab (-6+ c^\prime) \\
%-abc^\prime & -3b(1+ \frac{a}2+xa) + x a b c & 3b (1- \frac{a}2) \\
%ab(-6 + c^\prime) & 3b (1-\frac{a}2) & 1-3 b (1+\frac{a}2-xa) -x abc
%\end{array}
%\right).
%%
%\end{eqnarray}
%

In the tetrahedral coordinate,
the $f$ coupling is written when $f \propto Y_2+\lambda(Y_3+Y_4)$ (case
B1) as
\begin{equation}
f^{\rm tetra} \propto
 \left(
\begin{array}{ccc}
-\frac12\lambda & 0 & 0 \\
0 & \lambda & \sqrt{\frac32}\lambda \\
0 & \sqrt{\frac32}\lambda & 1
\end{array}
\right),
\label{caseB1}
\end{equation}
and
if the notation in the previous section is used,
we obtain
 $a:b:c = (\sqrt3 \cos\psi + \sqrt2 \sin\psi):(\sqrt2 \cos\psi - \sqrt3
\sin\psi):-1$
where $\tan2\psi = \sqrt6 \lambda/(1-\lambda)$.
The mass squared ratio
is $\Delta m^2_{\rm sol}/\Delta m^2_{\rm atm} = 3/4 \lambda^2
(1-4\lambda + O(\lambda^2))$.
When we use $f \propto Y_2+\lambda(Y_3+Y_5)$ (case B2),
\begin{equation}
f^{\rm tetra} \propto
 \left(
\begin{array}{ccc}
-\frac12\lambda & 0 & \frac{\sqrt3}2\lambda \\
0 & \lambda & 0 \\
\frac{\sqrt3}2\lambda & 0 & 1
\end{array}
\right),
\label{caseB2}
\end{equation}
we obtain
$a:b:c = (\sqrt3 \cos\psi^\prime - \sin\psi^\prime):\sqrt2:(-
\cos\psi^\prime - \sqrt3\sin\psi^\prime)$
where $\tan2\psi^\prime = \sqrt3 \lambda/(1+\lambda/2)$.
The mass squared ratio
is $\Delta m^2_{\rm sol}/\Delta m^2_{\rm atm} = 3/4 \lambda^2 (1-\lambda
+ O(\lambda^2))$.

We note that if there is a $O(\lambda)$ correction in the 12 element
in the tetrahedral coordinate,
it modifies the $\theta_s$ angle largely, and it separates from the
nearly tri-bimaximal mixing,
and thus we do not use the choice.

As we have obtained in the previous section,
$U_{e3}$ prediction is
\begin{equation}
|U_{e3}| \sim \left| \sqrt{\frac12 \frac{\Delta m^2_{\rm sol}}{\Delta
m^2_{\rm atm}}}
+ e^{i \gamma} \frac{1}{3\sqrt2} V_{us} \right|.
\label{eq-ue3}
\end{equation}
Clearly,
if the parameters $\rho$, $\lambda$, $\epsilon$ are all real,
then, $\gamma = 0$ or $\pi$,
and
the maximal and minimal values of
$U_{e3}$ are obtained.
At that time, there is no phase in the PMNS mixing matrix.
(The Kobayashi-Maskawa phase can be obtained from a phase of $r_2$
and/or $r_3$.)

In Fig.2 (case B1, Eq.(\ref{caseB1})) and Fig.3 (case B2,  Eq.(\ref{caseB2})),
we plot  $U_{e3}$ when $\rho$ $\lambda$, $\epsilon$ are real
to find the lower and upper limits.
Due to the off-diagonal elements of $f$ in the tetrahedral coordinate,
the atmospheric angle shifts from $45^{\rm o}$,
and the shift is correlated to $U_{e3}$, unlike the case of Model I.

In case B1,
using the experimental constraint on $\Delta m^2_{\rm
sol}/\Delta m^2_{\rm atm}$, we find that $U_{e3}$ is predicted to be 0.05-0.08.
This solution corresponds to the sign choice $e^{i\gamma} = -1$ in Eq.(\ref{eq-ue3}).
%
%where $\gamma$ is a phase which is different form the MNSP phase
%(but it may be correlated).
%%
%$U_{e3}$ can be larger than 0.1 in this case.
%Thus, if the center value of the MINOS result holds,
%we need to choose the case B.
%\newpage
%
We also plot atmospheric mixing angle in Fig.2 (right).
It is interesting to note that
for the case B1, %choice $f \propto Y_2+\lambda(Y_3+Y_4)$,
$\lambda$ should be negative for $|\lambda| \sim 0.1-0.3$
to fit mass squared difference ratio since
$\Delta m^2_{\rm sol}/\Delta m^2_{\rm atm} = 3/4 \lambda^2 (1-4\lambda + O(\lambda^2))$.
As a result, the direction of the shift is determined to fit
experimental values, i.e. $\theta_{\rm atm}> 45^{\rm o}$.

\begin{figure}
\centering
\mbox{\subfigure{\includegraphics[viewport = 10 15 280 220,width=3in]{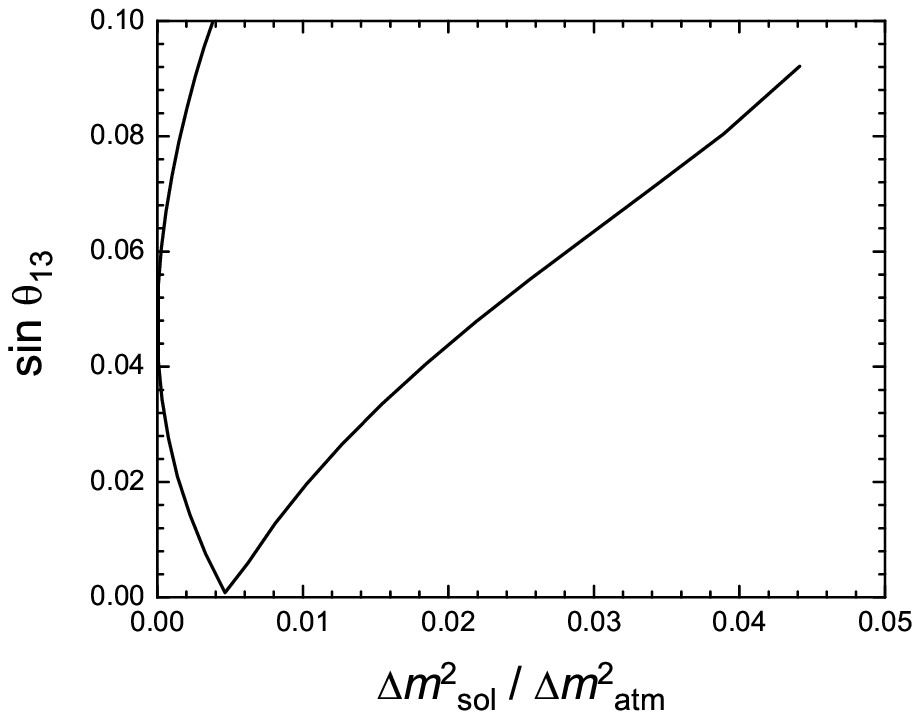}}\quad
\subfigure{\includegraphics[viewport = 10 15 280 220,width=3in]{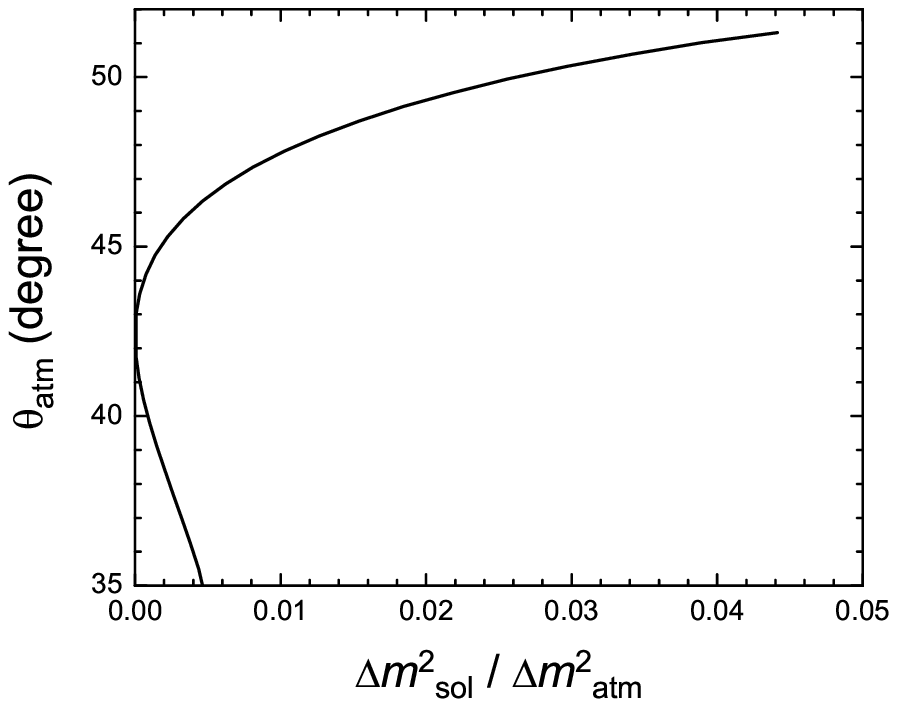} }}
\caption{$\sin\theta_{\rm 13}$ is shown as a function of $\Delta m^2_{\rm
sol}/\Delta m^2_{\rm atm}$ (left) and $\theta_{\rm atm}$ is shown as a function of $\Delta m^2_{\rm
sol}/\Delta m^2_{\rm atm}$ (right) for Model II-case B1 (described in the text).} \label{fig34}
\end{figure}

\begin{figure}
\centering
\mbox{\subfigure{\includegraphics[viewport = 10 15 280 220,width=3in]{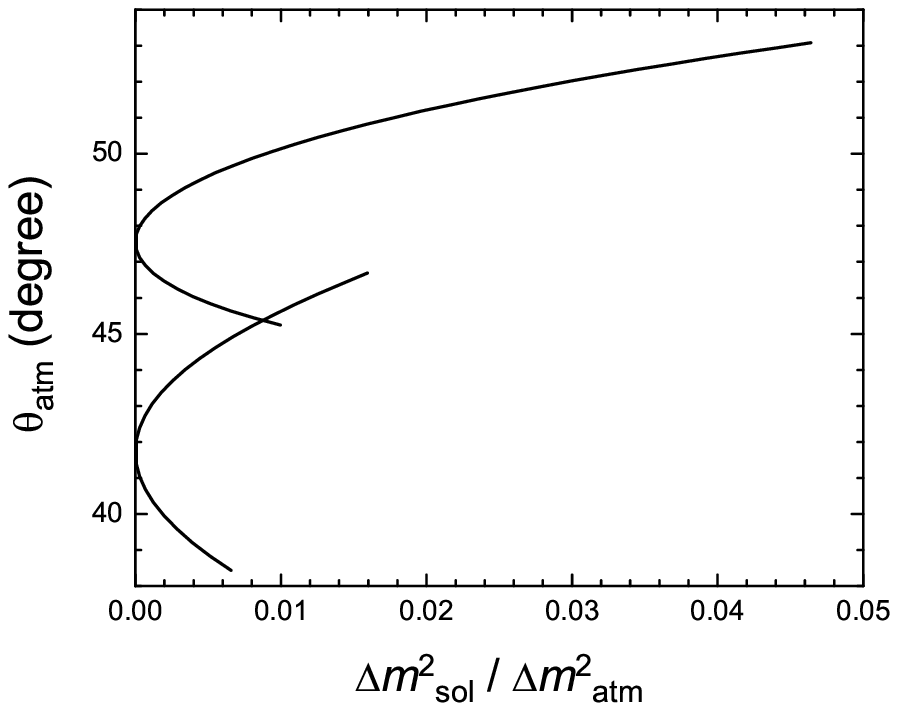}}\quad
\subfigure{\includegraphics[viewport = 10 15 280 220,width=3in]{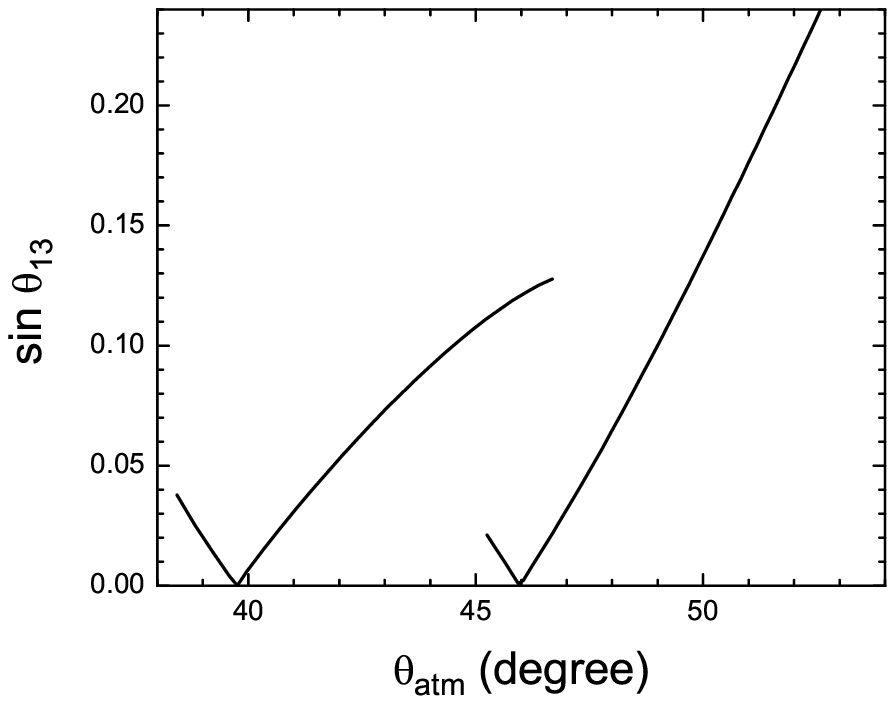} }}
\caption{$\theta_{\rm atm}$ is shown as a function of $\Delta m^2_{\rm sol}/\Delta m^2_{\rm atm}$ (left)
and $\sin\theta_{\rm 13}$ is
shown as a function of $\theta_{\rm atm}$ for Model II-case B2 (described in the text).} \label{fig56}
\end{figure}

\begin{figure}
\centering
\mbox{\subfigure{\includegraphics[viewport = 10 15 280 225,width=3in]{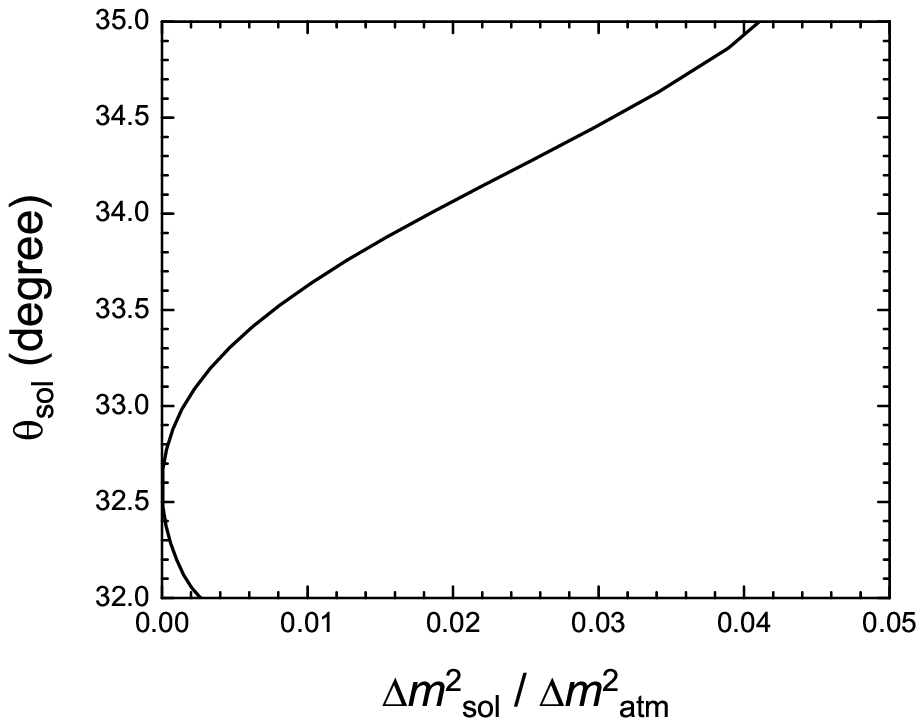}}\quad
\subfigure{\includegraphics[viewport = 10 15 280 225,width=3in]{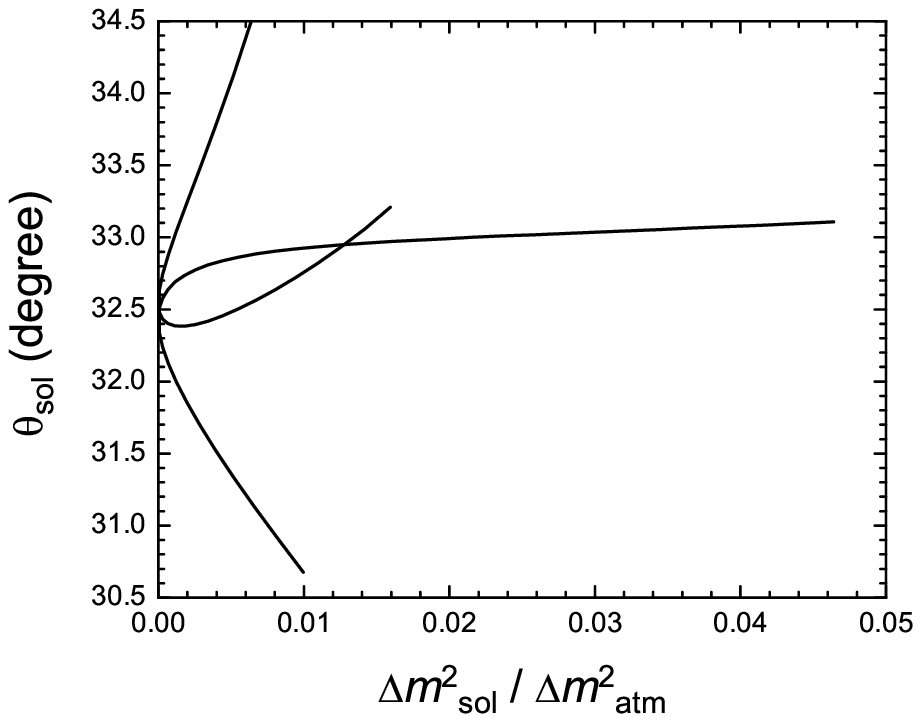} }}
\caption{$\theta_{\rm sol}$ is shown as a function of $\Delta m^2_{\rm
sol}/\Delta m^2_{\rm atm}$ for Model II-case B1 (left), case B2 (right).} \label{fig4}
\end{figure}

In case B2, %$f \propto Y_2+\lambda(Y_3+Y_5)$,
the mass squared ratio can be fitted for both signatures
of $\lambda$ (we find two branches in both graphs in Fig.3).
The plot in Fig.3 is shown constraining $\tan^2\theta_{\rm sol} > 0.35$,
and one can find that
 $\Delta m^2_{\rm sol}/\Delta m^2_{\rm atm}$ becomes too small for one of
the branch
to fit solar mixing angle,
and $\theta_{\rm atm}> 45^{\rm o}$ is favored in this case as well.
 Larger $U_{e3}$ values ($>0.15$) is preferred once we include the
experimental limit on
$\Delta m^2_{\rm sol}/\Delta m^2_{\rm atm}$.
 The solution corresponds to the sign choice $e^{i\gamma} = +1$ in
Eq.(\ref{eq-ue3}).

 As we discussed above that both case A and case B solutions can be
obtained in Models I and II.
 The interesting question is how can we distinguish these two models. For
example,
 the predictions of $U_{e3}$ (as shown in Figs 1,2 and 3) distinguishes
between the cases A and B.
Actually, if $U_{e3}$ is just below the current CHOOZ bound,
the case B solution with $e^{i\gamma}=1$ is preferred.
Since the case A is a more natural solution for Model I and case B is a
more natural solution for Model II,
one may weakly conclude that this prediction distinguishes between
Models I and II.
But a stronger way to distinguish  these models would be to use the
predictions of
$\theta_{\rm atm}$.
In Model I, $\theta_{\rm atm}$ is fixed to be 45$^{\rm o}$ up to $\pm
2-3^{\rm o}$ corrections from $V_{cb}$,
where as Model II prefers $\theta_{\rm atm} >45^{\rm o}$ (Figs 2 and 3)
in the experimentally allowed region.
This difference is directly due to the rigorousness of $\mu-\tau$ symmetry
in the $f$ coupling from the tri-bimaximal ansatz ($V_\nu=1$ (Model I))
where the $f$ coupling
has $Z_2\times Z_2$ symmetry.
In Model I, the deviation from the maximal angle is related to $V_{cb}$, while
in Model II, the deviation is related to $U_{e3}$.
The current best fit value of the atmospheric mixing angle
($\theta_{\rm atm}^{\rm best\, fit} = 43^{\rm o}$ \cite{Fogli:2008cx})
is nearly the maximal mixing,
and it implies the Model I.
%
%(best fit value is $\theta_{\rm atm} = 43^{\rm o}$ \cite{Fogli:2008cx}),
However, the error is still large.
The accurate deviation from the maximal angle
will be obtained in future three generation fit of the neutrino
oscillations \cite{GonzalezGarcia:2004cu},
and it will give us an important test for the tri-bimaximal ansatz.

The predictions for $\theta_{\rm sol}$
are similar with a small margin in these models,
since the deviations from the tri-bimaximal angle ($\theta_s = 35.3^{\rm o}$)
are related to $U_{e3}$ in all cases.
In Model I, $\theta_{\rm sol}$ is predicted to
be $\sim 32^{\rm o}$, where as, in Model II (Fig. 4),
$\theta_{\rm sol}$ is predicted to be $\sim 34^{\rm o}$ (case B1),
$\sim 33^{\rm o}$ (case B2),
once we include all the experimental bounds.
The PMNS phase is assumed to be 0 or $\pi$ in the plot,
and the general phase fit will change the predictions of the angle,
especially for Model II.

\section{Derivation of rank one ansatz}

The rank one Yukawa coupling with {\bf 10} Higgs field
generates the features of flavor hierarchy,
and rank 1 matrices can often appear in various ways (flavor symmetry,
discrete symmetry,
and string models).
In this section, we give an
SO(10) model, where the rank one ansatz used in our discussion of flavor
emerges from a discrete symmetry.

When the direct couplings of chiral fermions with a Higgs field
are forbidden by a symmetry,
and the effective Yukawa couplings are generated by propagating
vector-like matter fields,
the rank of the effective Yukawa matrix depends on the number of the
vector-like fields.
Actually, when there are only one pair of vector-like matter fields
as a flavor singlet, the effective Yukawa matrix is rank 1.

The model we assume has one extra
vector-like pair of matter fields with mass slightly above the GUT scale
(denoted by
$\psi_V\equiv {\bf 16}_V\oplus \bar\psi_V\equiv \overline{\bf 16}_V$)
and three gauge singlet
fields $Y_a$. We add a $Z_4$ discrete symmetry to the model under which
the fields $\psi_a\rightarrow i \psi_a$, and $Y_a\rightarrow -iY_a$.
The {\bf 10}-Higgs field $H$
%
%$H\equiv {\bf 10}$-Higgs,
%$\Delta\equiv {\bf 126}\,\oplus \,\bar\Delta\equiv \overline{\bf 126}$-Higgs
is invariant under this symmetry.
The gauge invariant Yukawa superpotential under this assumption
is given by
\begin{eqnarray}
W~=~\psi_V H \lambda\psi_V~+
M_V\psi_V\bar{\psi}_V~+~\bar{\psi}_V\sum_a Y_a\psi_a.
\end{eqnarray}
When we give vevs $\langle Y_a\rangle\neq 0$,
$\psi_V$ and $\psi_a$ are mixed.
The heavy vector-like fields, $\bar\psi_V$ and a linear combination of
$\psi_V$ and $\psi_a$ (i.e. $M_V \psi_V+ \sum_a Y_a \psi_a$),
and
the effective operator below its
scale and at the GUT scale is
given by:
\begin{eqnarray}
{\cal L}_{eff}~=~\frac{\lambda}{M_V^2+ \sum_a Y_a^2}\left[\sum_aY_a\psi_a\right] H \left[\sum_b Y_b\psi_b\right].
\end{eqnarray}
This gives rise to a rank one $h$ coupling.
We note that it does not contradict the $O(1)$ top Yukawa coupling,
when $M_V^2 \sim \sum_a Y_a^2$ (or $M_V^2 < \sum_a Y_a^2$).

If we let the $\overline{\bf 126}$ Higgs field
transform like $-1$ under $Z_4$, it can induce the $f$ coupling with
rank three.

Another way to get mass matrix patterns of types in sec. IV is to
assume that there are three component flavon fields (denoted by
the dimensionless field $\phi_i\equiv \frac{\Phi}{M}$) which are
representations of some internal flavor group and constrain their
couplings to fermions by other symmetries. As an example, let us
choose four flavon fields which transform as follows under a
$Z_4\times Z_2\times Z_2$ group: $\phi_1 (-i, +,+)$; $\phi_2
(1,-,+)$; $\phi_3(1, +, -)$; $\phi_4(1,-,-)$; $\psi(i,+,+)$;
$\bar{\Delta}(-1, +,+)$; $H(1,+,+)$; $H^\prime(-1,-,-)$. The
invariant Yukawa coupling under these symmetries is:
\begin{eqnarray}
L_Y~=~\phi_1^t\phi_1\psi\psi H+\phi_2^t\phi_2\psi\psi\bar\Delta
+\phi_3^t\phi_3\psi\psi\bar\Delta +\phi_4^t\phi_4\psi\psi\bar\Delta~+~
\phi_2^t\phi_3\psi\psi H^\prime~+~h.c.
\end{eqnarray}
In general the vevs of the $\phi_i$ fields align as in Eq. (25) which can
then lead to our type of rank one models. With suitable discrete
symmetries, e.g., %$Z_7 \rtimes Z_3$ and
$\Delta (27)$ \cite{Luhn:2007sy}, the flavon vevs can align as in
Eq. (26), leading to models of the type considered here (Model I).
When $Z_2$ group is chosen instead of $Z_2\times Z_2$ group such
that $\phi_1 (+)$; $\phi_2 (-)$; $\phi_3(-)$; $\phi_4(+)$;
$\psi(+)$; $\bar{\Delta}(+)$; $H(+)$; $H^\prime(+)$,
$\phi_2^t\phi_3$ term is allowed in the $\psi\psi\bar\Delta$
coupling and Model II (B1) can be considered. More details on the
flavon vev alignment with discrete symmetries and implications for
rank one models is currently under investigation.

\section{Conclusion}
In conclusion, we have shown how a simple ansatz for the dominant Yukawa
coupling matrix in renormalizable SO(10) models can lead to a unified
understanding of the diverse quark and lepton flavor hierarchies. We
suggest this as a possible way to address the challenge of a unified
description of quark-lepton flavor. We have not attempted in this note to
derive our ansatz from any specific discrete symmetries, although we
show a guideline to obtain a rank one form from a higher than GUT
scale theory. This may be the next step towards a complete theory of
flavor.

Within our rank one hypothesis, we have considered two classes of
models which are nearly tri-bimaximal and point out that both a
measurement of $\theta_{13}$ and the atmospheric mixing angle
$\theta_{\rm atm}$ can distinguish between these models. In these
models, the natural minimal value of $\theta_{13}$ is around 0.05,
whereas the maximal value can be larger than 0.15. This range of
$\theta_{13}$ can be probed in the upcoming experiments. The
current estimate of $\theta_{13}$ using the 1.5$\,\sigma$ excess
of events in MINOS $\nu_\mu-\nu_e$ appearance channel~\cite{minos}
and all other experimental data  is $\sin^2\theta_{13}\simeq
0.02\pm 0.01$ (1$\,\sigma$)~\cite{Fogli:2009ce}.

\section*{Acknowledgement}
The work of R.~N.~M. and Y. M. is supported by the US National
Science Foundation under grant No. PHY-0652363  and that of B. D.
is supported in part by the DOE grant DE-FG02-95ER40917.


\begin{thebibliography}{99}
\bibitem{footnote}
A $3\times 3$ unitary matrix $U$ can be expressed
as $U = P_1 \bar U P_2$
where
$P_1$ and $P_2$ are diagonal phase matrices
and $\bar U$ is a standard form of the unitary matrix
which includes one phase and three mixings.
The usual expression of the CKM matrix is parameterized in the $\bar U$ part
since $P_1$ and $P_2$ are unphysical.
%
The phase matrix
${\rm diag}(1,
e^{i\alpha}, e^{i\beta})$ corresponds to the $P_2$ part of $\tilde V_e$
which is physical in the PMNS matrix.
Actually, one can find that the phase $\alpha$ corresponds to the PMNS phase
(which appears in the long baseline neutrino oscillation)
approximately up to a quadrant.
\end{thebibliography}
\end{document}